 \documentclass[final,3p,times,twocolumn]{elsarticle}

\usepackage{amssymb}

\usepackage{graphicx}
\usepackage{dcolumn}
\usepackage{bm}
\usepackage{xcolor}
\usepackage{changes}
\usepackage[normalem]{ulem}
\usepackage[colorlinks=true,
            linkcolor=blue,
            urlcolor=blue,
            citecolor=blue]{hyperref}
\journal{Computational Materials Science}
\newcommand{\hb}[1]{\textcolor{black}{#1}}
\newcommand{\fl}[1]{\textcolor{black}{#1}}

\newcommand{\etal}{\textit{at al.}}
\definechangesauthor[name=Markus Aichhorn, color=red]{MA}
\definechangesauthor[name=Florian Lindner, color=blue]{fl}
\usepackage{booktabs} 
\usepackage{xcolor,colortbl} 
\begin{document}

\begin{frontmatter}

\title{Decoupling the effects of geometry and nature of strain in LaMnO$_3$: Interplay of dynamic correlations and uniaxial strain driving magnetic phase transitions}

\author[TU Graz]{Florian P Lindner}
\author[TU Graz]{Markus Aichhorn}
\affiliation[TU Graz]{organization={Institute of Theoretical and Computational Physics, Graz University of Technology},
            addressline={Petersgasse 16/III}, 
            city={Graz},
            postcode={8010}, 
            state={Styria},
            country={Austria}}

\author[Cambridge,Dundee]{Hrishit Banerjee\corref{cor1}}
\affiliation[Cambridge]{organization={Yusuf Hamied Department of Chemistry, University of Cambridge},
addressline={Lensfield Road}, city={Cambridge}, postcode={CB2 1EW}, state={Cambridgeshire}, country={United Kingdom}}
\affiliation[Dundee]{organization={School of Science and Engineering, University of Dundee},
            addressline={Nethergate}, 
            city={Dundee},
            postcode={DD1 4HN}, 
            state={Scotland},
            country={United Kingdom}}
\ead{hb595@cam.ac.uk, hbanerjee001@dundee.ac.uk}
\cortext[cor1]{Corresponding author}

\begin{abstract}
Recent years have seen tremendous progress in experimental techniques to create uniaxial strain. Motivated by these advances 
we investigate the effect of uniaxial strain on LaMnO$_3$  employing ab-initio dynamical mean-field theory, and put it in contrast to biaxial strain that occurs in epitaxial systems.
 Projecting on the low-energy subspace of Mn $3d$ states, and solving multi-impurity problems, our approach emphasizes on local dynamic correlations at Mn sites. At ambient pressures, LaMnO$_3$ crystallizes in an orthorhombic unit cell, with in-plane lattice constants $a<b$, and shows an A-type antiferromagnetic ground state. If we apply uniaxial compressive strain such that the in-plane lattice becomes square with lattice constant $a$, we find a ferromagnetic insulating state. 
 This is in sharp contrast to DFT results using various functionals like PBE, PBE+$U$, and hybrid functionals like HSE, which all predict a half-metallic ferromagnetic behaviour.
 Interestingly, applying uniaxial tensile strain, such that the in-plane lattice becomes square with the longer lattice constant $b$, an antiferromagnetic insulating state is observed. We trace back these results to the reduction in Jahn-Teller distortion in the case of compressive strain, favoring a ferromagnetic state. This reduction is absent in the tensile case, and the antiferromagnetic state therefore survives. Our study shows that it is the flavour of the strain (compressive or tensile) which is decisive for the magnitude of Jahn-Teller distortions and, hence, the magnetic state. 
\end{abstract}




\end{frontmatter}



\section{Introduction}
The effect of strain in oxide systems, which gives rise to exotic states not seen in bulk, has been 
of particular interest for quite some time to the condensed matter community. 
Strain tuning is particularly useful in optimising properties to suit the needs of device applications. It is effective in tuning both material properties of use in practical applications and also in tuning electronic correlations and magnetism which is interesting from a theoretical standpoint thereof. 

Hwang \etal~\cite{HWANG2019100} have shown that when complex oxides are engineered as thin films, their chemical and physical properties can be modified to be markedly different from their bulk form, providing additional degrees of freedom in materials design. They explored the landscape of strain-induced design of complex oxides in the context of oxygen electrocatalysis and ferroelectricity. Strain-driven modification of metal-oxygen bond length and octahedral distortion in perovskites has been shown to influence oxide electronic properties. 
These strain studies have been performed along with the advances in state-of-the-art thin-film fabrication and characterization that have enabled a high degree of experimental control in realizing strain effects in oxide thin-film systems. 
For ferroelectric properties, strain engineering can both enhance polarization in known ferroelectrics and induce ferroelectricity in material systems that would not be ferroelectric in bulk otherwise.

Franchini~\etal~\cite{franchini} showed that epitaxial strain offers an effective route to tune the physical parameters in transition metal oxides. The effects of strain on the bandwidths and crystal-field splitting have long since been known, however recent experimental and theoretical works have shown that the effective Coulomb interaction changes as well upon structural modifications. This particular effect has a huge influence on current material engineering studies based on epitaxy-based material synthesis. In their theoretical study, they show that the response \hb{ of correlation to applied} strain is strongly dependent on the material for different oxides with a different occupation of the $d$ shell. Studying LaTiO$_3(d^1)$, LaVO$_3(d^2)$, and LaCrO$_3(d^3)$, correlation effects are shown to be significantly altered by the effect of strain on oxides, however, further investigations may be warranted.

There are several ways of generating strain in oxides. The most common is applying a biaxial strain by growing one lattice on another epitaxially in the form of a heterostructure through methods like Molecular Beam Epitaxy (MBE) or Pulsed Laser Decomposition (PLD). This of course leads to the famous family of oxide heterostructures~\cite{Ohtomo2004}, which also generates the polar 2-dimensional electron gas (2DEG). However, modern piezoelectric methods of strain generation, recently developed by Hicks \etal~\cite{hicks}, also enable one to apply uniaxial strain on crystals and films. This particular method is quite intriguing since it enables the study of the effects of strain without dealing with the polar 2DEG.

LaMnO$_3$ (LMO) exhibits a very diverse phase diagram in the bulk state in both doped and undoped phases~\cite{satpathy, zhang}. In undoped bulk LMO, orbital order due to Jahn-Teller (JT) distortions of the MnO$_6$ octahedra sets in at reasonably high temperatures of around $750$\,K. Subsequently, magnetic exchange between Mn$^{3+}$  ions leads to the formation of A-type antiferromagnetic (A-AFM) insulating phase with N\'{e}el temperature of approximately $140$\,K in which ferromagnetic (FM) planes are coupled in an antiferromagnetic manner.   \textcolor{black}{The origin of the cooperative Jahn-Teller distortion and orbital order in LaMnO$_3$ has also been carefully investigated using many-body methods like DFT+DMFT. The interplay of Kugel-Khomskii superexchange with lattice distortions and their effect on transition temperatures have been calculated, showing the importance of dynamic correlations \cite{pavarini2010}.} The primary characteristic of LMO is that the stoichiometric compound can be doped by cautiously controlling its stoichiometry. Doping La by Ca or tuning oxygen content, a phase
transition can be triggered in bulk LMO from an AFM insulating state to a FM metallic state. On the other hand, such a doping effect can also be a major disadvantage in identifying \hb{the chemical composition of} a system. Hence tuning of magnetic and electronic properties in oxides by strain is nowadays emerging as one of the key methods of tunability. 

Investigations using density functional theory (DFT) supplemented with Hubbard $U$ (DFT+$U$) on epitaxially strained LMO, with strain corresponding to that of an STO substrate, showed (primarily in the context of studying LMO/STO heterostructures) \cite{Lee_2013_PAPER} a suppression of the JT distortion and a ferromagnetic ground state. This state is, however, metallic rather than insulating, as claimed in LMO/STO experiments. 
In a further DFT+$U$ study of LMO strained to STO~\cite{gong}, the structural relaxation allowing for symmetry lowering to monoclinic structure, and resulting antiferro orbital ordering between symmetry inequivalent Mn atoms
was used to explain the ferromagnetic insulating behavior of LMO/STO. However, such symmetry lowering may be difficult to be accommodated within a heterostructure geometry, where LMO is constrained between the layers of cubic STO from both top and bottom and unable to deviate from the cubic 
symmetry. \textcolor{black}{This is further corroborated by experimental X-ray diffraction studies of LMO/STO superlattices which clearly show a tetragonal lattice \cite{lmo-sto-expt}. This contradicts the prediction of a monoclinic structure by structure prediction methods. It also indicates that the hypothesis of orbital order due to monoclinicity is not valid.}  

In yet another study using hybrid functionals, it was found that applying $\sim 6$\,GPa of uniaxial pressure along the [010] direction activates the transition to a ferromagnetic half-metallic state  \cite{Rivero_2016_PAPERa}.

\textcolor{black}{There have been several claims of local symmetry breaking giving rise to paramagnetic insulating states in DFT, without invoking +U corrections of any type. Standard DFT calculations based on single unit cells have been labelled "naive" DFT \cite{zunger1, zunger2, Varignon2019}. The premise here has been that if one uses a large enough supercell, one can easily break the symmetry locally to get paramagnetic states and insulating behaviour. 
This has naturally been applied to well-known materials like LMO. What has been completely ignored in this endeavor is the fact that the paramagnetic states in these materials are dependent on temperature, whereas DFT is a ground state $T=0$\,K theory even in the non "naive" versions. Moreover, the DOS presented in some of the work does not show symmetrical behaviour between the spin channels and cannot claim to be paramagnetic \cite{Varignon2019}. In any case, the application of compressive strain for example has the property of reducing JT distortions drastically \cite{banerjee2019}, and hence it is imperative to identify the interplay of strain and correlations in giving rise to exotic electronic states like ferromagnetic insulating behaviour arising in correlated oxides, as well as its evolution to higher temperatures and paramagnetic phases.}

In a previous study~\cite{banerjee2019}, Banerjee \etal{} have shown the effect of \hb{ biaxial epitaxial} strain on LaMnO$_3$ \hb{however from a single particle DFT picture using both DFT+$U$ and hybrid functionals}. It was shown that compressive epitaxial strain exerted by epitaxial matching of LMO to a square substrate significantly reduces the JT distortion in LMO, which in turn favours a ferromagnetic (FM) ground state instead of the antiferromagnetic (AFM) state in bulk LMO. 
\hb{However the relationship between dynamic correlation and strain remained largely unexplored and hence it was important to investigate the interplay of strain and dynamic correlation using a many-body method. In a subsequent study}~\cite{banerjee-mplb, banerjee2020} it was also shown that this transition can also be driven by strong correlations from a dynamical mean-field theory (DMFT) based many-body perspective. For the ``strained-bulk'' structure generated by applying strain on the bulk structure of LMO 
it was found that DFT+DMFT yields a ferromagnetic insulating solution for small enough temperature. The Curie temperature is found to be $\sim 100$\,K. 

The effect of uniaxial strain on LMO and its interplay with strong correlation remains unknown. While hybrid functionals as used in Ref.~\cite{Rivero_2016_PAPERa} do have a better description of short-range non-local exchange, the treatment of correlations is still at the level of DFT, which as shown in the case of bi-axial epitaxial strain does not capture the effect of strong correlations. In this study, we employ ab-initio DMFT calculations to capture the effects of correlation on an uniaxially strained LMO. We study two specific cases. First, 
compressive strain, where the longer in-plane lattice parameter is set equal to the shorter in-plane lattice parameter, and second tensile strain, where the shorter in-plane lattice parameter is set equal to the longer, thereby creating a square in-plane lattice in both cases. This allows us to disentangle not only the effects of compressive or tensile strain but also the impact of the square lattice \textcolor{black}{symmetry itself, which has not yet been addressed in studies with epitaxially generated biaxial strain.} 

\textcolor{black}{We find that on the application of a uniaxial compressive strain such that the in-plane lattice becomes square with the smaller lattice constant $a$, a ferromagnetic insulating state emerges. This is in stark contrast to DFT results predicting a half-metallic ferromagnetic behaviour in this case, due to the static treatment of correlations \cite{Rivero_2016_PAPERa}.
On the other hand, if we produce a square lattice with the larger lattice constant $b$, the ground state is antiferromagnetic just as the unstrained sample is. Our study clearly shows that it is the flavour of the strain (compressive or tensile) which has a major effect on the JT distortions, and not the effect of geometry/symmetry of matching to a square lattice, which is impossible to determine based on application of biaxial strain generated epitaxially. We also show the evolution of the magnetic states to higher temperatures and their corresponding paramagnetic phases. The study of uniaxial strain thus is crucial to differentiate between \textcolor{black}{two effects, namely a reduction of JT distortions and an increase in lattice symmetry.} Our study, thus, helps to understand possible tunabilities of magnetic states by uniaxial strain.}

\section{Computational Details \label{Computational Details}}

For the DFT geometry relaxations, we used the Vienna Ab-initio Simulation Package (VASP)~\cite{Kresse_1993_PAPER, Kresse_1996_PAPER, Kresse_1996_PAPERa}. VASP uses a plane-wave basis with projector-augmented wave (PAW) potentials~\cite{Kresse_1999_PAPER}. Throughout all relaxations, the cutoff-energy for the plane-wave basis was set to 560\,eV, and a \textcolor{black}{$7\times7\times7$} Monkhorst-Pack~\cite{Monkhorst_1976_PAPER} k-point-mesh, along with a GGA exchange-correlation functional in the parametrization of Perdew-Burke-Ernzerhof (PBE)~\cite{PBE_1996_PAPER} were used. The cutoff energies of the PAW potentials were 219\,eV for La (11 valence electrons, 5s2 5p6 5d1 6s2), 270\,eV for Mn (7 valence electrons, d6 s1) and 400\,eV for O (6 valence electrons, s2 p4).

The effects of strong correlations between the Mn-$d$ electrons of LMO beyond single-particle DFT were accounted for within a DFT+DMFT framework. The input for such calculation schemes is usually converged non-magnetic DFT results. Non-magnetic DFT input was calculated by WIEN2k~\cite{Blaha_2020_PAPER}, which is a Linearized Augmented Plane Wave + Local Orbital (LAPW+LO) code. For the non-magnetic self-consistent field calculations done in WIEN2k, 
a shifted k-mesh with 2500 k-points was used. The parameters $R\cdot K_\textrm{max}$ and $G_\textrm{max}$ were set to 7.0 and 12.5 respectively. \textcolor{black}{For the calculations in WIEN2k we used a GGA functional in PBE parametrization as well.} The resulting Mn $d$ band-structure from these non-magnetic calculations was found to be metallic. Due to cubic crystal field splitting the Mn $d$-states are split into $t_{2g}$ and $e_{g}$ manifolds.

To carry out DFT+DMFT calculations, an effective low-energy Hamiltonian was constructed from a basis set of projective Wannier functions~\cite{Aichhorn_2009_PAPER} for the Mn-$d$ states. The projective Wannier functions in this work were constructed using the TRIQS/DFTTools~\cite{Aichhorn_2016_PAPER} package. The full Mn-$d$ manifold, including the $t_{2g}$ as well as $e_{g}$ states, were selected to be treated as correlated, to account for the possibility of high-spin states. The energy window $\mathcal{W}$ for the projection was set to be $\mathcal{W} = [-1.38, 2.86]$\,eV relative to the Fermi energy. The interaction Hamiltonian $H_{int}$ was selected to be a Slater-Type Hamiltonian in density-density approximation.
 ~For the interaction parameters $U$ and $J_H$ we used 6.0\,eV and 0.75\,eV, respectively. These are rather standard literature values for DFT+DMFT calculations for manganites~\cite{banerjee2019, banerjee2020}. \textcolor{black}{If not explicitly stated otherwise, for all DFT+DMFT calculations the inverse temperature $\beta$ was set to 80\,eV\textsuperscript{-1}}. 

To find magnetic solutions, spin splitting was introduced in the real parts of the self-energies in the first DMFT iteration. \fl{This is equivalent to providing an initial guess for the magnetic moment when doing a spin-polarized DFT calculation. The purpose of the introduced small spin split in the self-energies is to make the calculation converge faster to a spin-polarized solution at lower temperatures. Furthermore, it prevents the calculation from running into meta-stable paramagnetic solutions.} The self-consistency cycle was performed using \textcolor{black}{the package TRIQS/DFTTools~\cite{Aichhorn_2016_PAPER,Parcollet_2015_PAPER}}. The resulting Anderson impurity problems were solved using a continuous-time quantum Monte Carlo solver in hybridization expansion~\cite{werner1,Gull_2011_PAPER}. We use two implementations of this solver, the matrix formulation as provided in the TRIQS/CTHYB package~\cite{Seth_2016_PAPER}, as well as a segment-picture version of this solver~\cite{ct-seg}. 


As off-diagonal elements cause severe sign problems in continuous time quantum Monte Carlo solvers, the impurity problems in this work were solved in a basis in which the local Wannier Hamiltonian is diagonal. 
The impurity problem to be solved consists of 10 (due to spin) orbitals, which we refer to as $A_{j}^{\sigma}$. Here $j \in \left\{0\cdots 4\right\}$ refers to the orbital index and $\sigma$ denotes the spin. All DFT+DMFT calculations in this paper are single-shot calculations, and a fully localized limit double-counting correction was used~\cite{Anisimov_1993_PAPER}. 
Real frequency spectral functions were obtained using maximum-entropy analytic continuation, as implemented in the TRIQS/MaxEnt package~\cite{maxent}.

 \begin{figure}[th]
\centering
\includegraphics[width=0.45\textwidth]{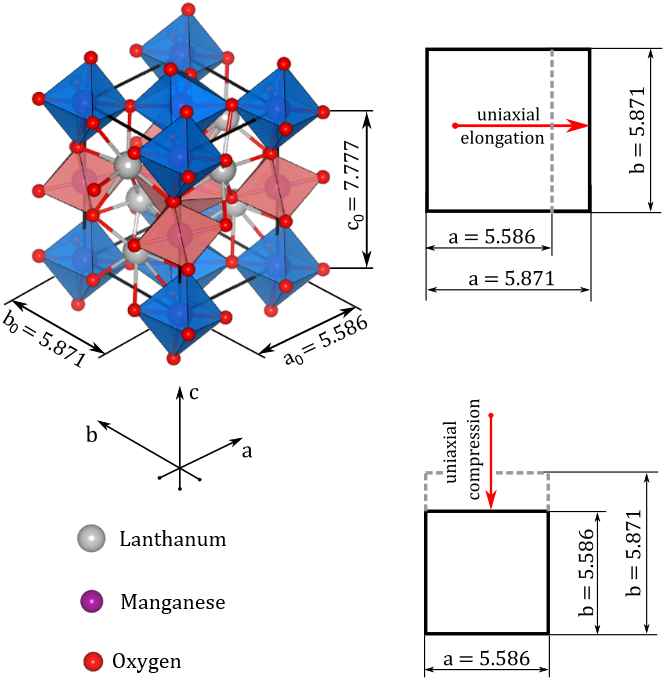}
\caption{Crystal structure of unstrained bulk LMO and corresponding ways of generating uniaxial strain. In its ground state, bulk LMO crystallizes in an orthorhombic crystal structure with Pnma space group symmetry. This is necessary in order to accommodate the Jahn-Teller (JT) distorted tilted and rotated MnO\textsubscript{6} octahedra, which form the basic building block of LMO. This structure was taken from the Materials Project (MP) database. The MP structure was relaxed in VASP (ionic relaxation). The top right panel shows the application of uniaxial tensile strain and the bottom right panel shows the application of uniaxial compressive strain.} \label{fig:crystal_cell_paper_new.png} 
\end{figure}        

\section{Results}

\subsection{Crystal Structure \label{Crystal Structure}}
We use an 
orthorhombic LMO unit-cell, containing four formula units as a starting point for our calculations. The used crystal structure has Pnma space group symmetry and was obtained from the materials project (MP) database \cite{Jain_2013_PAPER}. The structure ID in the MP database is 17554, which is the lowest energy structure on the convex hull for stoichiometric LaMnO$_3$ composition, and has an orthorhombic space group \cite{LMO_MP_CELL}.
The structure \textcolor{black}{given in MP as per MP documentation} was relaxed using a GGA+$U$ calculation performed in VASP, where an implementation of DFT+$U$ as of Dudarev \textit{et al.} was selected \cite{Dudarev_1998_PAPER}. Furthermore, according to the MP documentation, $U=3.9$\,eV was used on the Mn sites, and a spin-polarised structural optimization was performed with a ferromagnetic initial moment. During the relaxation, all three degrees of freedom-the ionic positions within the cell, the cell volume itself and the cell shape-were allowed to change. According to the MP database, the resulting ground-state structure is half-metallic and in a ferromagnetically ordered state. The MP database structure, along with the lattice parameters is shown in the left part of Fig.~\ref{fig:crystal_cell_paper_new.png}. We visualized the crystal structure using VESTA \cite{Momma_2011_PAPER}.    

\subsubsection{Unstrained LMO \label{Unstrained LMO}}
\begin{table*}
\caption {\fl{Overview of important structural parameters of the LMO crystal structures used throughout this work. The lattice constants for unstrained orthorhombic (Pnma) LMO were taken from \cite{Jain_2013_PAPER}, and internal coordinates were obtained from an A-AFM spin-polarized relaxation of the atomic positions with VASP. The cells for the compressive and the tensile strained structures were obtained from BM fits, to FM and A-AFM spin-polarized $(E,V)$ datapoints, respectively, that were also produced using VASP. The reported experimental values were measured at 4.2\,K, see \cite{Jain_2013_PAPER} for further details.\\}}  
\label{tab:LMO_structural_parameters_exp_strained} 
\centering
\begin{tabular}{ccccc}
\toprule 
 & \textbf{experiment} & \textbf{unstrained} & \textbf{compressive} & \textbf{tensile}\\
\midrule
\rowcolor{black!20} \textbf{lattice parameters [\AA]} & \vphantom{1} & \vphantom{1} & \vphantom{1} & \vphantom{1}\\ 
\rowcolor{black!20} a & 5.532  & 5.586 & 5.586 & 5.871\\ 
\rowcolor{black!20} b & 5.742  & 5.871 & 5.586 & 5.871\\
\rowcolor{black!20} c & 7.668  & 7.777 & 7.873 & 7.504\\
\textbf{orthorhombicity [\%]} & \vphantom{1} & \vphantom{1} & \vphantom{1} & \vphantom{1} \\
$\left(\frac{b}{a}-1,\frac{c}{a\sqrt{2}}-1 \right)$& (3.8 / -2.0)  & (5.1 /-1.6) & (0.0 / -0.3) & (0.0 / -9.6)\\
\rowcolor{black!20}\textbf{volume per f. u. [$\AA^{3}$]} & 60.89 & 63.76 & 61.42 & 64.66\\
\textbf{Mn-O} & \vphantom{1}&\vphantom{1}\\
\textbf{bond length [\AA]} & \vphantom{1}&\vphantom{1}&\vphantom{1}&\vphantom{1}\\
l & 2.187 & 2.220 & 2.025 & 2.310\\
m & 1.959 & 2.008 & 2.015 & 1.950\\
s & 1.905 & 1.956 & 2.010 & 1.954\\  
\rowcolor{black!20}\textbf{JT modes in [\AA]} & \vphantom{1} & \vphantom{1}& \vphantom{1} & \vphantom{1}\\
\rowcolor{black!20}as in (\ref{Q1_and_Q2_in_LMO}) & \vphantom{1} & \vphantom{1} &\vphantom{1}&\vphantom{1}\\
\rowcolor{black!20} $Q_{2}$ & 0.399 & 0.401 & 0.021 & 0.503\\
\rowcolor{black!20} $Q_{3}$ & -0.142 & -0.139 & -0.004 &-0.297\\
\textbf{applied uniaxial strain [\%]} & - & - & -4.85 & 5.10 \\
\bottomrule
\end{tabular}
\end{table*}
     
The ground-state electronic configuration of the Mn\textsuperscript{3+} transition metal ion in LMO is $t_{2g}^{\uparrow\uparrow\uparrow}e_{g}^{\uparrow}$ (high spin). The singly occupied $e_{g}$ sub-level is subject to a strong cooperative Jahn-Teller (JT) effect, which causes the MnO\textsubscript{6} octahedra to distort, rotate in the ab-plane, and tilt along the $c$ direction, as shown in Fig.~\ref{fig:crystal_cell_paper_new.png}. This effect, which is of crucial importance to understanding the electronic and magnetic properties of LMO, is characterized by long (l) and short (s) length Mn-O bonds in the $ab$-plane, and medium (m) length Mn-O bonds along the $c$-direction. The most important JT modes that cause the mentioned cooperative distortions are the in-plane mode $Q_{2}$ and the out-of-plane mode $Q_{3}$~\cite{Rivero_2016_PAPER,Lee_2013_PAPER}. In terms of the Mn-O bond lengths these modes are given by   
\begin{equation}
    Q_{2} = \frac{2(l-s)}{\sqrt{2}} \\ Q_{3} = \frac{2(2m-l-s)}{\sqrt{6}}
\label{Q1_and_Q2_in_LMO}
\end{equation}

As a first step, we carry out an A-AFM and an FM spin-polarized relaxation of the ionic positions on the MP structure just described, using VASP.
From this, we find that the FM optimized structure is lower in energy by only 6\,meV per f.u. \fl{This - at first glance - contradictory prediction of our VASP calculation and the well-established A-AFM ground state from experimental studies, is most likely owed to the pre-bias of the materials project relaxation, which was done using GGA+U and FM spin-polarization. \textcolor{black}{This procedure of relaxation} reduces the lattice constants and leads to a smaller volume unit cell than what is needed for a DFT A-AFM ground state. 
In order to understand the origin of this discrepancy we use 
the AFM relaxed structure from hereon in our VASP calculation and push our calculations towards convergence, using  
a 16x16x16 Monkhorst-Pack, k-point mesh and an energy cut-off for the plane wave basis set of 600 eV. 
We find that 
the A-AFM ordered magnetic state is now lower in energy, separated from the FM ordered state by 1.3\,meV per f.u.} From subsequent calculations for FM and A-AFM order in WIEN2k on the \fl{same A-AFM relaxed structure}, we find \fl{again} the A-AFM ordered state to be \fl{the ground state of the unstrained LMO structure. Wien2k gives the AFM state separated from the FM state by an energy difference of 9\,meV per f.u.}
 \fl{Moreover in this WIEN2k calculations, we used stricter settings than for the non-magnetic calculations, reported in section \ref{Computational Details}. Namely, we used a shifted k-mesh with 3200 k-points and parameters $R\cdot K_\textrm{max}$ and $G_\textrm{max}$ of 8.25 and 12.75 respectively. 
As a result, both codes provided the use of strict convergence criteria to predict the A-AFM ordered state to be the ground state. The remaining difference, -1.3 vs. -9\,meV, has to be considered as within the DFT accuracy of our calculations.
From this analysis, we conclude that there is strong competition between the FM and A-AFM ordered states for the unstrained LMO unit cell.}

 Based on these observations and in agreement with experimental results, we decided to use the A-AFM relaxed crystal structure from VASP, for our further considerations. Thus, our unstrained LMO unit cell has in-plane lattice constants $a_{0}=5.586$\,\AA, $b_{0}=5.871$\,\AA, and a $c$-axis constant of $c_{0}=7.777$\,\AA. The unstrained unit cell has a volume of 255.05\,\AA\textsuperscript{3} and  we find the two JT modes $Q_{2}$ and $Q_{3}$ to equal 0.401\,\AA{} and -0.139\,\AA, respectively. From these values we see that the unstrained structure is strongly JT distorted.   

Table~\ref{tab:LMO_structural_parameters_exp_strained} summarizes the most important structural parameters of our unstrained bulk LMO unit cell \fl{along with experimental lattice constants and JT modes, reported in \cite{ELEMANS1971238}. The calculated energy differences between the two competing magnetic phases within the unstrained bulk LMO crystal structure are reported in table \ref{tab:LMO_electronic_properties_unstrained_strained}. 
We have shown above that the discrepancy in the magnetic ordering for the unstrained system is due to computational accuracy, which we may estimate to be $\pm 5$\,meV. This implies that our conclusions for the magnetic ground states of the strained systems: FM for compressive and A-AFM for tensile-are robust and would not change with stricter and more costly computational settings.}


\begin{table*}
\caption {\fl{Overview of electronic properties for the unstrained and strained bulk LMO cells, with structural parameters as reported in table \ref{tab:LMO_structural_parameters_exp_strained}. The stabilized magnetic order is reported in brackets next to the denoted values. All VASP calculations have been done using the settings given in section \ref{Computational Details}. The discrepancy in the magnetic order in the unstrained case is due to computational accuracy, see text for more details.}}  
\label{tab:LMO_electronic_properties_unstrained_strained} 
\centering
\begin{tabular}{cccc}
\toprule 
 & \textbf{unstrained} & \textbf{compressive} & \textbf{tensile} \\
\midrule
\rowcolor{black!20}\textbf{Energy difference between} & \vphantom{1}& \vphantom{1} & \vphantom{1}\\
\rowcolor{black!20}\textbf{FM and AFM phase [meV / f.u.]}  & \vphantom{1} &\vphantom{1}&\vphantom{1}\\
\rowcolor{black!20} $\Delta E$(VASP) & -6 (FM) & -46 (FM) & 20 (AFM)\\
\rowcolor{black!20} $\Delta E$(WIEN2k)  & 9 (AFM) & -54 (FM) & 32 (AFM)\\
\textbf{DFT+DMFT Band-Gap [eV]}  & \vphantom{1} & \vphantom{1} & \vphantom{1} \\
\textbf{Paramagnetic Phase}  & 3.0 & 2.0 & 3.5\\
\textbf{Magnetic Phase}  & 2.8 (AFM) & 1.4 (FM) & 3.4 (AFM)\\
\bottomrule
\end{tabular}
\end{table*}

\subsubsection{Uniaxially strained LMO}
Next, we explore the influence of uniaxial strain on the crystal structure of LMO, especially the tunability of the JT modes $Q_{2}$ and $Q_{3}$ by this sort of strain. As the magnetic and electronic properties of LMO depend strongly on the JT distortions within the compound, we also want to look at the possibility of using uniaxial applied strains to induce magnetic phase transitions in LMO. In order to mimic the effects of uniaxial strains we perform "strained bulk" calculations.

\paragraph{Compressive Strain \label{Compressive Strain}}
We study the effect of uniaxial compressive strain by setting the longer in-plane lattice constant $b_{0}=5.871$\,\AA{} of the unstrained unit cell to the exact same value as $a_{0}$, thus forming an in-plane square lattice with constant $a=b=a_0=5.586$\,\AA. This imitates the process of \fl{applying -4.85 \% of} uniaxial compressive strain, to compress the longer in-plane lattice parameter of the unstrained crystal structure to match the shorter in-plane lattice parameter. A schematic visualization of this process is shown in the bottom right corner of Fig.~\ref{fig:crystal_cell_paper_new.png}. We allow the $c$-axis parameter to relax, and optimize $c$ (as well as the ionic positions) of the \hb{resultant tetragonal (due to $a=b=a_0$)} perovskite unit. The relaxation is done under the constraint that the in-plane lattice constants $a$ and $b$ remain fixed. We obtain the optimal $c$-axis parameter of 7.873\,\AA{} for our compressively strained structure by fitting $(V,E)$-data points for 10 different values of $c$ to a Birch-Murnaghan (BM) equation of state. In order to produce the required data points we performed spin-polarized relaxations in VASP \textcolor{black}{with computational parameters as} described in section \ref{Computational Details}. The two considered magnetic orders were FM and A-AFM. For our calculations we found the FM ordered structure to be lowest in energy for all considered c values. Hence we calculated the optimal c-lattice parameter for the uniaxial \textcolor{black}{compressed} structure from the BM optimal volume of the ferromagnetic BM parabola, which we find to be 245.65\,\AA\textsuperscript{3}. The minima of the A-AFM and the FM BM parabolas, obtained as just described, are separated by an energy difference of 46\,meV per f.u., and thus we conclude that the FM phase is preferred in the compressed unit cell, in contrast to the unstrained cell where small energy difference between FM and AFM phases indicated strong competition between magnetic phases, see section \ref{Unstrained LMO}. \fl{For the so-constructed BM optimal compressed structure also WIEN2k energy calculations for an FM and A-AFM phase were performed. We find an FM groundstate 54\,meV per f.u. below the A-AFM state. The calculations were done using the strict convergence criteria reported in section \ref{Unstrained LMO}. } Furthermore, we observe a dramatic reduction of JT modes $Q_{2}$ and $Q_{3}$ in the compressed cell. We tabulate the most important structural parameters of the obtained compressed LMO unit cell in table~\ref{tab:LMO_structural_parameters_exp_strained}. Our observation of the dramatically reduced JT modes for compressive uniaxial strain also matches several previous findings~\cite{Rivero_2016_PAPER, Rivero_2016_PAPERa, Lee_2013_PAPER, banerjee2019, banerjee2020}. Rivero~\textit{et al.}~\cite{Rivero_2016_PAPER} in particular showed that the JT modes are very sensitive towards uniaxial strains. Finally, we point out that due to the compressive strain a structural phase transition has occurred, namely from the cooperatively Jahn-Teller distorted, orthorhombic Pnma cell, to a more symmetric undistorted pseudo-cubic cell.

\paragraph{Tensile Strain \label{Tensile Strain}}
Now we want to clarify if this reduction in JT distortion is due to the higher symmetry of the strained structure \hb{generated by matching the xy plane to a square lattice by putting $a=b=a_0$} or, if it is really the flavor of the applied strain \hb{(compressive/tensile)} that is of importance. We proceed as \textcolor{black}{before} and imitate \fl{the application of 5.1 \% of} uniaxial tensile strain, acting on the orthorhombic LMO unit cell from sub-section \ref{Unstrained LMO} by setting the lattice parameter $a_{0}$ of the former to the exact same value as $b_{0}$. Hence, we again carry out "strained bulk" calculations on a structure with an in-plane square lattice, however, this time with a larger constant of $a=b=b_0=5.871$\,\AA. We visualize the process of uniaxial tensile strain elongating the $a$ axis to match $b_0$ in a schematic in the top right corner of Fig.~\ref{fig:crystal_cell_paper_new.png}. Allowing the ionic positions to relax, we determine the optimal $c$-axis parameter of 7.504\,\AA~ from the minima of BM parabolas, which we again obtain from fits to the data from FM and A-AFM spin-polarized VASP relaxations. Interestingly we observe that in the case of the tensile strained structure, the BM parabola for the A-AFM data points is lower in energy for all considered $c$ values, with the minima of the FM and A-AFM BM parabolas being separated by 20\,meV per f.u. Therefore we calculate the optimal $c$-axis parameter for the tensile-strained crystal structure from the optimal BM volume obtained from the A-AFM BM parabola, which we find to be 258.65\,\AA\textsuperscript{3}. The structural parameters of the optimized tensile-strained crystal structure are also listed in table \ref{tab:LMO_structural_parameters_exp_strained}. \fl{As in the case of uniaxial compressive strain, WIEN2k energy calculations with tight parameter settings as reported in section \ref{Unstrained LMO}, predict the A-AFM magnetic phase in the obtained tensile strained structure to be lower in energy by 32\,meV per f.u. than the FM ordered state. So also in the tensile strained case, we find a perfect agreement between the predicted ground states of the two DFT codes.} The second very interesting observation for the tensile strained unit cell is that we find the JT distortions within this structure to behave very differently as compared to the compressed crystal structure. In the latter, we find $Q_{2}$ and $Q_{3}$ effectively reducing to negligibly small values, whereas in the tensile case we observe (compared to the unstrained cell) an \textit{increase} of both $Q_2$ and $Q_3$ modes by 25 \% and 113\%, respectively. This is also in accordance with the indicated stabilization of the A-AFM ordered phase that we found when determining the optimal $c$-axis constant from the BM fits. 
\hb{Similar to} the unstrained orthorhombic LMO unit cell the JT distortions lift the two-fold degeneracy of the Mn-d sublevels $e_{g}:=\left\{d_{z^{2}},d_{x^{2}-y^{2}}\right\}$. Due to the large magnitude of the in-plane mode $Q_{2}$ one finds a pronounced anti-ferro orbital order within the Mn planes. According to the Goodenough-Kanamori rules this leads to ferromagnetic in-plane magnetic order. The coupling between the layers originates from antiferro super-exchange between the Mn-$t_{2g}$ orbitals and results in an overall A-AFM order. Contrary to the case of the uniaxial compressed structure from paragraph \ref{Compressive Strain} we find no indication of a structural or a magnetic phase transition in the structure formed by uniaxial tensile strain. Hence we conclude that it is really the flavor of the applied stain that is important for strain-tuning the magnetic properties of LMO.

\subsection{Electronic and magnetic structure in DFT+DMFT}

An observation of the JT modes $Q_{2}$ and $Q_{3}$ reveals that uniaxial compressive strain causes a structural phase transition, whereas for tensile strain no such transition was observed, see table \ref{tab:LMO_structural_parameters_exp_strained}. In the following we calculate the electronic and magnetic properties of the uniaxial strained LMO cells using DFT+DMFT. Compared to DFT approaches as in Ref.~\cite{Rivero_2016_PAPERa}, in our calculations we include local dynamic correlation on the Mn sites in order to account for strong correlation effects between the Mn-$d$ electrons. 

\subsubsection{Unstrained LMO \label{Unstrained LMO electronic structure}}

\begin{figure}[h]
\centering
\includegraphics[width=0.46\textwidth]{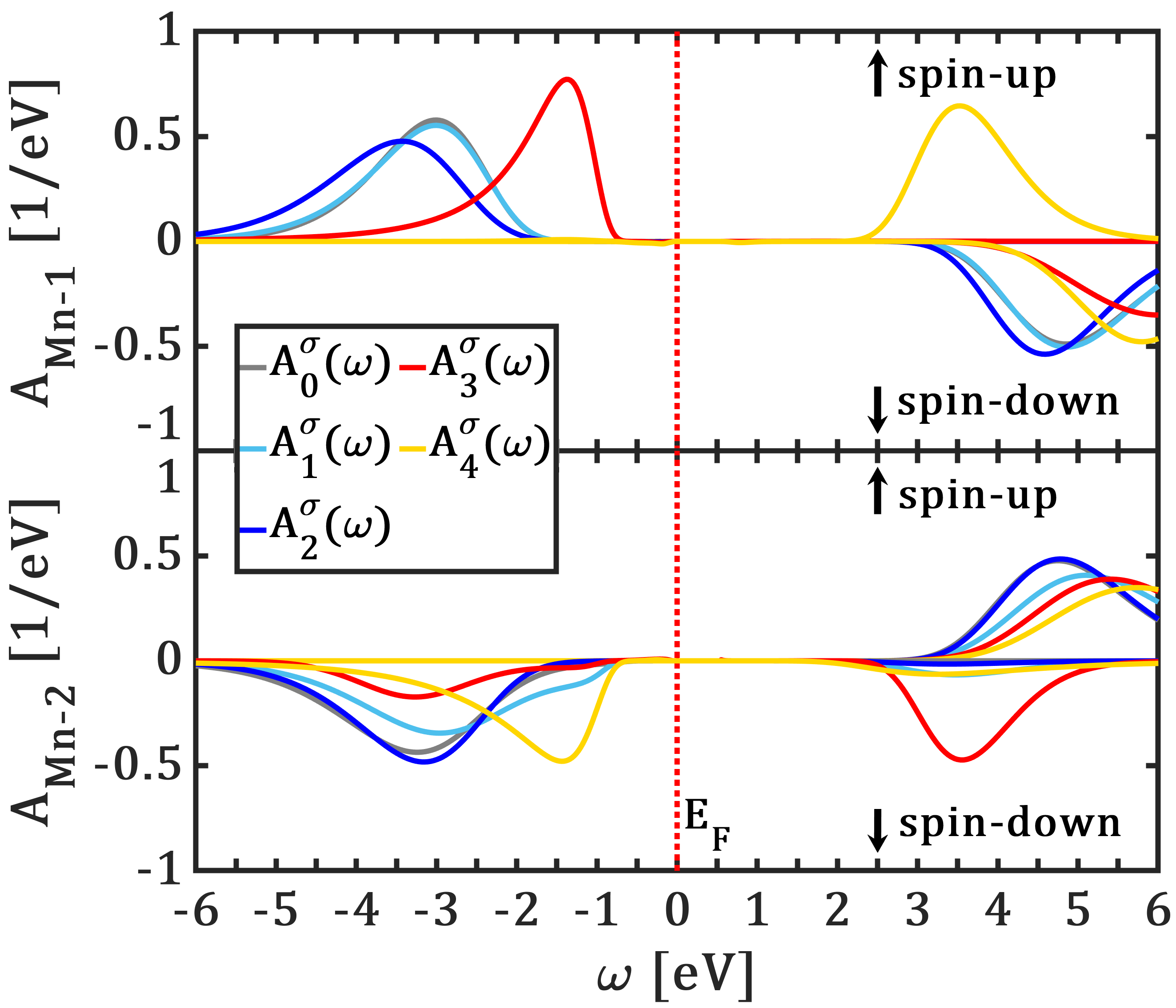}
\caption{Orbitally- and spin-resolved DFT+DMFT spectral functions for unstrained LMO in the A-AFM order. Upper panel: First inequivalent Mn site. Lower panel: Second inequivalent Mn site. Calculations are done for $\beta=80$\,eV$^{-1}$, $U=6.0$\,eV, and $J_H=0.75$\,eV.}
\label{fig:spectral_function_unstrained_bulk_LMO.png}
\end{figure} 

From experiments~\cite{Elemans_1971_PAPER} and theoretical studies~\cite{Rivero_2016_PAPER, banerjee2019} it is known that the ground state of LMO is insulating and of A-AFM magnetic order. This order, in which ferromagnetic ordered Mn layers are coupled anti-ferromagnetically along the $c$ direction is shown in Fig.~\ref{fig:crystal_cell_paper_new.png}. 
As explained in section~\ref{Unstrained LMO}, using solely DFT calculations \hb{at} the level of GGA functionals, it is rather difficult to arrive at a conclusion on the ground state of the unstrained LMO unit cell. \hb{Hence we resort to DMFT calculations for the correct description of electronic structure.} 
We use the A-AFM relaxed structure from VASP to prepare the non-magnetic DFT input for our DFT+DMFT calculations. As a first step we tested for a FM ordered solution in this structure by doing a single-impurity spin-polarized DFT+DMFT calculation. This means that all four Mn sites in the unit cell are treated as structurally and magnetically equivalent and that in the first DMFT iteration, we introduce a small spin-splitting to induce FM order. \textcolor{black}{After each iteration, we calculate the Wannier moments. 
This calculation does not lead to a stable self-consistent FM solution. 
Instead, when the moments were pointing upwards in iteration $n$, they point downwards in iteration $n+1$ with similar size, and vice versa. This is indicative that the system tends to moment formation and magnetic ordering, but that the ordering pattern does not fit into the chosen unit cell~\cite{georges_rmp, banerjee2022, banerjee2023}. For our case at hand it means that the system tends to AFM ordering. We want to note that the calculation does not converge to a paramagnetic ground state.}

After having ruled out an FM ordered solution within the unstrained unit cell we moved on to exploring the experimentally observed A-AFM ordered solution in DFT+DMFT. For this purpose, we expanded the single-impurity unit cell in order to accommodate the A-AFM order. The now eight Mn sites in the enlarged unit cell are grouped into two inequivalent Mn impurity sites. Fig.~\ref{fig:crystal_cell_paper_new.png} shows the two types of Mn sites as blue and red octahedra, respectively. The two impurities have opposite spin orientations.      


\begin{figure}[h]
\centering
\includegraphics[width=0.46\textwidth]{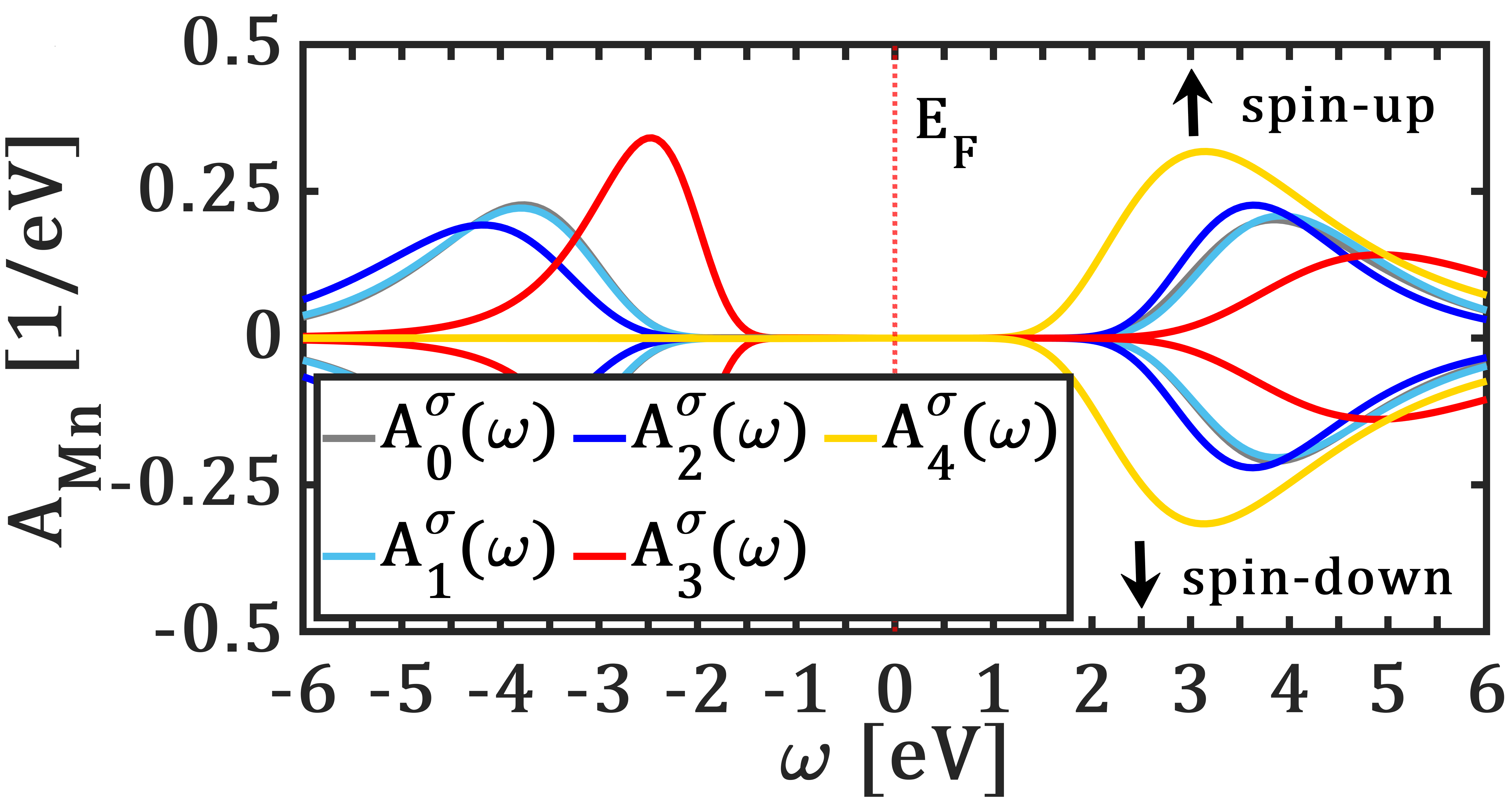}
\caption{Orbitally- and spin-resolved DFT+DMFT spectral functions for unstrained LMO in the paramagnetic phase. Calculations are done for $\beta=20$\,eV$^{-1}$, $U=6.0$\,eV, and $J_H=0.75$\,eV.}
\label{fig2a}
\end{figure} 

After introducing an \textcolor{black}{anti-ferro} spin splitting on the Mn sites in the very first DMFT iteration, we found that the DMFT loop converged to an insulating solution with an average A-AFM Wannier moment of 3.9\,$\mu_{B}$. 

The resulting spectral function for both impurities is shown in Fig.~\ref{fig:spectral_function_unstrained_bulk_LMO.png}. 
We observe a \hb{single-impurity} gap of approximately 2.8\,eV.   
%
\textcolor{black}{The spectral functions in figure 2 represent the 5 $d$ orbitals of Mn and show the Mn ions to be in $d^4$ high-spin state, because  
there are three half-filled $t_{2g}$ orbitals, one half filled $e_g$ orbital and one empty $e_g$ orbital. }

\textcolor{black}{The paramagnetic spectral function is shown in Fig. \ref{fig2a}. It has been calculated above the Neel temperature at $\beta=20$\,eV$^{-1}$ and, hence, does not show any magnetic ordering. The band gap is about $\sim 3$\,eV, which is of similar size as in the magnetic solution, see above. This shows that the system is an insulator at all temperatures.}

\subsubsection{Uniaxially strained LMO}

Next we apply DFT+DMFT calculations to obtain the electronic and magnetic structure of the uniaxially strained LMO unit cells. In order for the obtained results to be comparable, we used the exact same parameters for the calculations as before.

\paragraph{Compressive Strain}
\begin{figure}[h]
\centering
\includegraphics[width=0.46\textwidth]{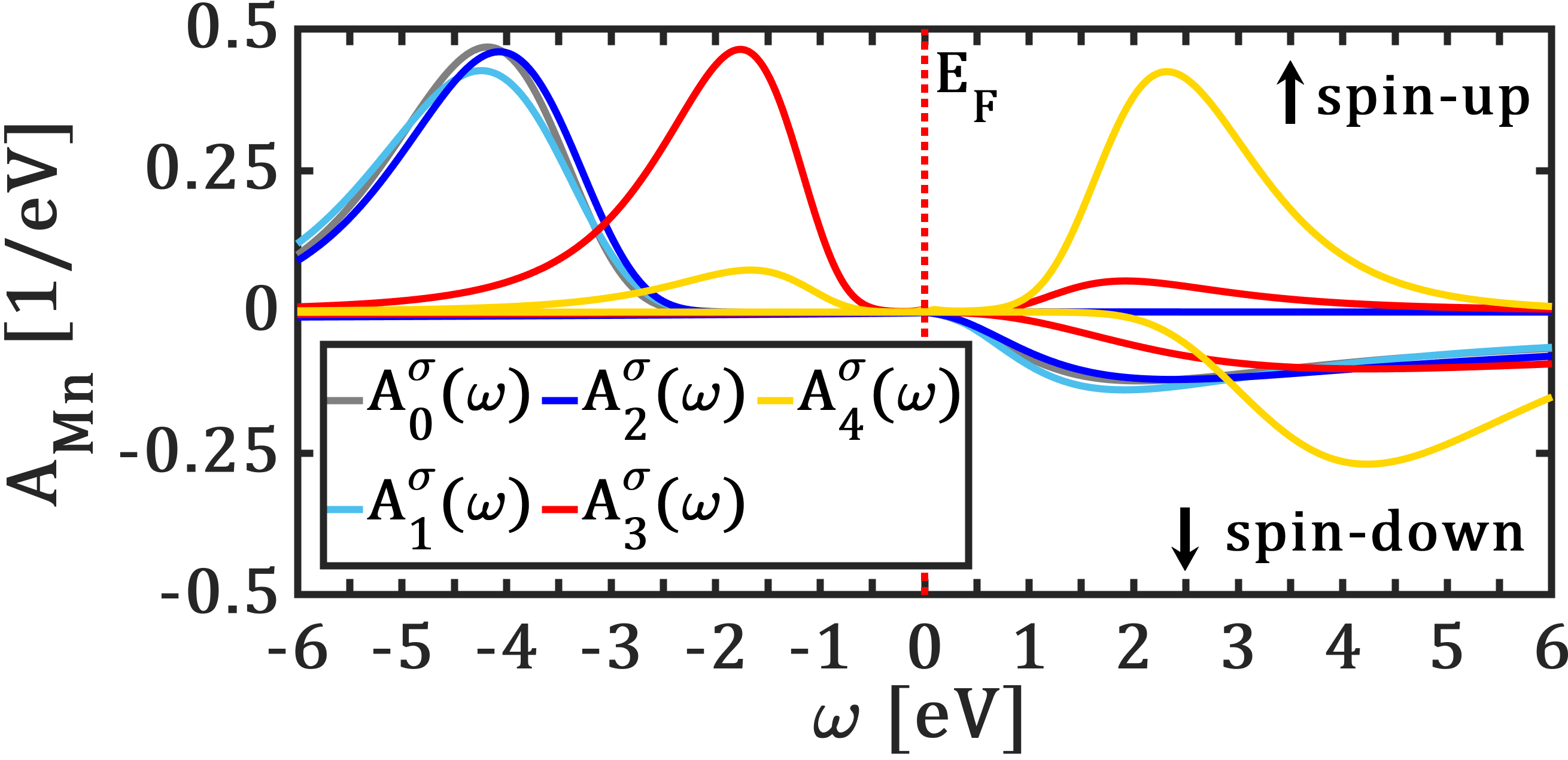}
\caption{Orbitally- and spin-resolved DFT+DMFT spectral functions for LMO under compressive strain. Calculations are done for $\beta=80$\,eV$^{-1}$, $U=6.0$\,eV, and $J_H=0.75$\,eV. The spectra are clearly ferromagnetic and insulating.}\label{fig:spectral_function_FM_compressive_strain.png}
\end{figure} 
 For the uniaxially compressed LMO unit cell we start again with a single-impurity DFT+DMFT calculation. Contrary to the single-impurity calculation on unstrained orthorhombic LMO, in this case, we find a stable self-consistent solution, with the Wannier moment of the impurity converged to 3.8\,$\mu_{B}$. This indicates that the system is in a high-spin ferromagnetic state.


In the next step, we analyze the obtained DFT+DMFT solution, by computing the impurity spectral function $A^{\sigma}_{orb}(\omega)$ for each spin channel and each of the five $d$ orbitals in the five-band low-energy model. The corresponding spectral function is shown in Fig.~\ref{fig:spectral_function_FM_compressive_strain.png}. $A^{\sigma}_{orb}(\omega)$ immediately reveals that the obtained DFT+DMFT solution describes a high-spin state with four half-filled orbitals in the majority spin channel. \textcolor{black}{These are three $t_{2g}$ orbitals and one $e_g$ orbital with a very small contribution from another $e_g$ orbital. The Mn is still in $d^4$ state. } The most interesting observation is that the inclusion of local dynamic correlation on the Mn sites opens up a gap of approximately 1.4\,eV also in the majority spin channel. Thus we conclude that the DFT+DMFT solution describes the uniaxial compressed LMO cell to be in an insulating FM ordered state. The insulating behavior of the FM ordered state is in contrast to the results reported by previous DFT studies~\cite{Rivero_2016_PAPERa}.
\begin{figure}[h]
\centering
\includegraphics[width=0.46\textwidth]{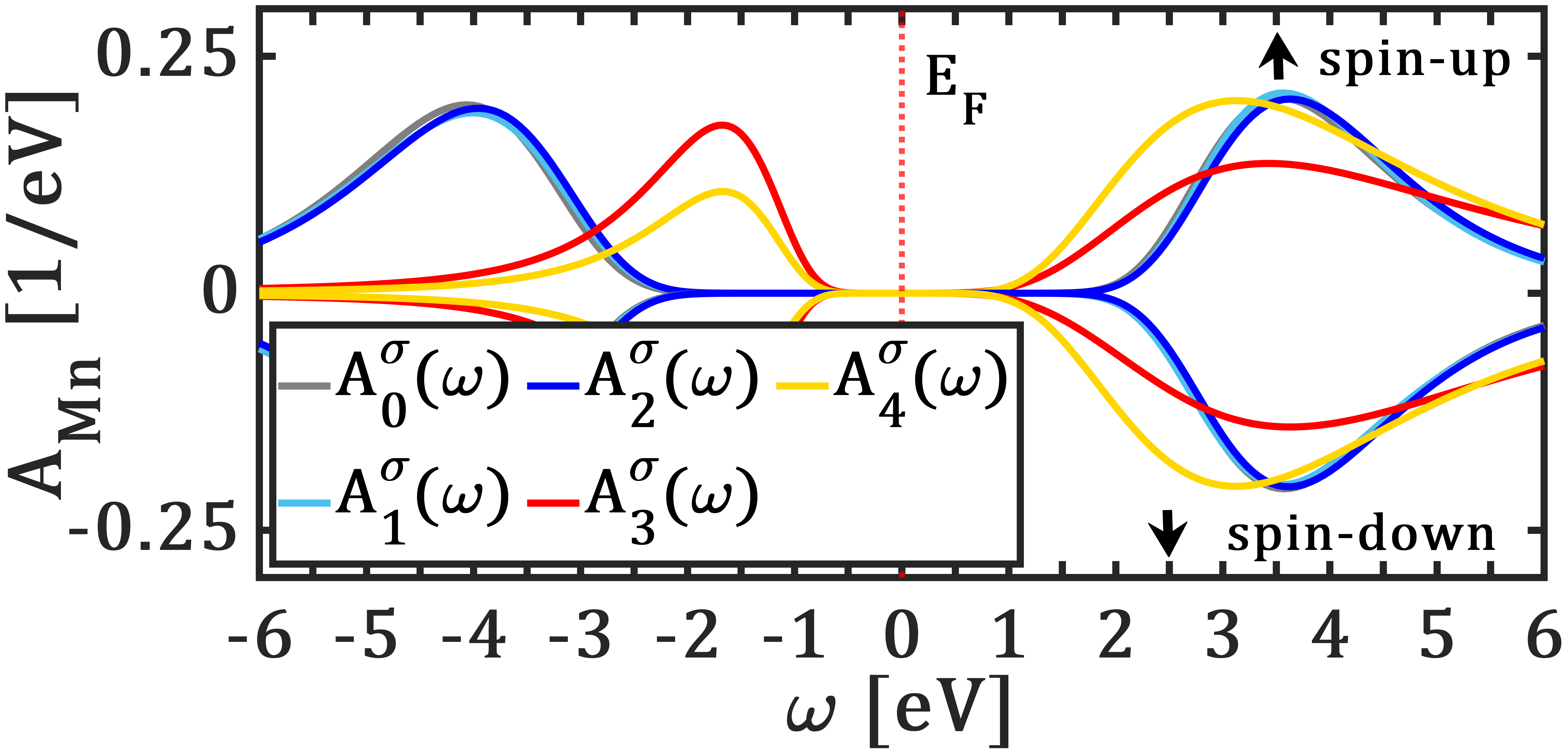}
\caption{Orbitally- and spin-resolved DFT+DMFT spectral functions for LMO under compressive strain in the paramagnetic phase. Calculations are done for $\beta=20$\,eV$^{-1}$, $U=6.0$\,eV, and $J_H=0.75$\,eV.}
\label{fig3a}
\end{figure} 


\textcolor{black}{The paramagnetic spectral function is shown in Fig. \ref{fig3a}. The band gap here is $\sim 2$\,eV, which is larger than in the FM solution above. In a sense, the FM solution tends to be closer to a metallic phase but does not go beyond the insulator-to-metal transition.}



\paragraph{Tensile Strain}
\begin{figure}[h]
\centering
\includegraphics[width=0.46\textwidth]{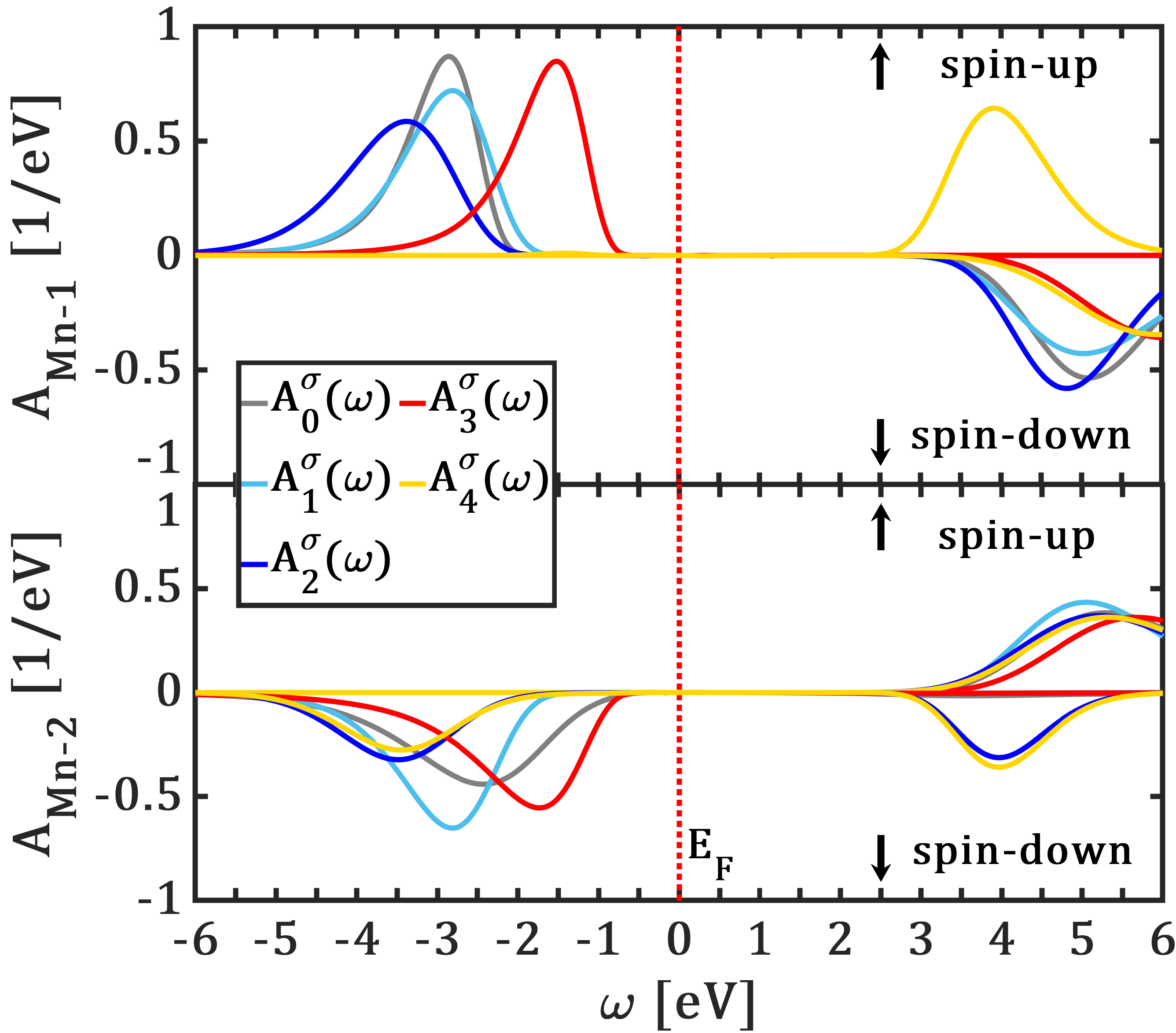}
\caption{Orbitally- and spin-resolved DFT+DMFT spectral functions for LMO under tensile strain, in the A-AFM order. Upper panel: First inequivalent Mn site. Lower panel: Second inequivalent Mn site. Calculations are done for $\beta=80$\,eV$^{-1}$, $U=6.0$\,eV, and $J_H=0.75$\,eV.}
\label{fig:spectral_function_uniaxial_tensile_strain_NEW.png}
\end{figure} 

Next, we investigate the case of tensile strain, from a DFT+DMFT perspective. Single particle DFT predicts an antiferromagnetic ground state for this phase. In order to confirm this we set up FM single-impurity DMFT calculations, in analogy to the unstrained case in section~\ref{Unstrained LMO electronic structure}. Also here, the calculations did not converge to a self-consistent solution, and we can rule out a FM ordered solution in the case of the uniaxial tensile strained unit cell. 
From these observations, we expect an A-AFM ordered solution in DFT+DMFT for the tensile-strained LMO unit cell. Going to the two-impurity unit cell it was indeed possible to stabilize such a solution. The two impurities were again formed by groups of two in-plane Mn sites, as shown in Fig.~\ref{fig:crystal_cell_paper_new.png}. Similar to the two-impurity calculation for the unstrained orthorhombic unit cell, we observe the Wannier moments for both impurities converging to a value of 3.9\,$\mu_{B}$. We further analyzed the solution by calculating the impurity spectral function for each of the five  $d$ orbitals and both impurities, see Fig.~\ref{fig:spectral_function_uniaxial_tensile_strain_NEW.png}.
\textcolor{black}{The spectral functions in this figure again represent the 5 $d$ orbitals of Mn and shows Mn to be in $d^4$ state for both impurities. In case of both impurities, there are three half-filled $t_{2g}$ orbitals, one half-filled $e_g$ orbital and one empty $e_g$ orbital.} 

From the spectral functions, we see that the insulating A-AFM solution exhibits a gap of approximately 3.4\,eV, which is 0.6\,eV larger than the observed 2.8\,eV gap in the unstrained case. Hence, we found that tensile strain does not induce a magnetic and/or structural phase transition in orthorhombic LMO. 
Furthermore, our result shows that the $Q_{2}$ and $Q_{3}$ breathing modes of the JT distortions in LMO are affected much more by the type of strain rather than the geometric effect of the square in-plane lattice constants. 
This finding underlines that the flavor of the applied strains, i.e, compressive versus tensile, is of great importance to induce magnetic phase transitions and, hence, corresponding strain tuning of magnetism in bulk LMO.          
\begin{figure}[h]
\centering
\includegraphics[width=0.46\textwidth]{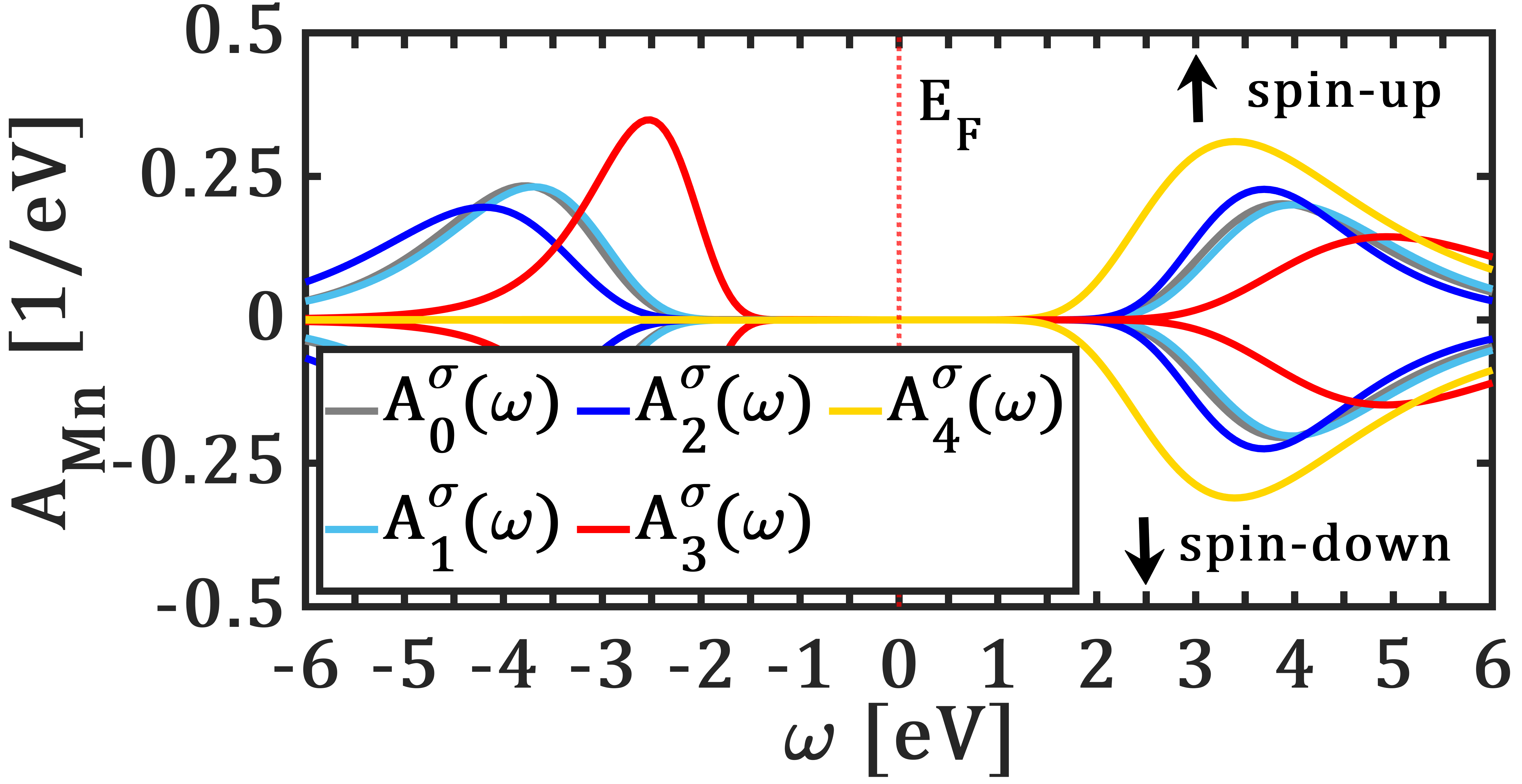}
\caption{Orbitally- and spin-resolved DFT+DMFT spectral functions for LMO under tensile strain in the paramagnetic phase. Calculations are done for $\beta=20$\,eV$^{-1}$, $U=6.0$\,eV, and $J_H=0.75$\,eV.}
\label{fig4a}
\end{figure} 

\textcolor{black}{The paramagnetic spectral function is shown in Fig. \ref{fig4a}. We find a gap of $\sim 3.5$\,eV, which is close to the gap in the AFM solution. In addition, the gaps found here in the tensile strained cases are very similar to the unstrained case presented in the beginning of the results section. We conclude that it is indeed the magnitude of the JT distortions that governs a large part of the physics of this compound.} 

\section{Summary and Conclusion}
In summary, we have shown the effect of uniaxial strain on LMO and its strong interplay with correlations. \fl{In the unstrained bulk LMO structure, DFT calculations indicate a strong competition between the A-AFM ordered, experimentally observed ground state, and a FM ordered phase. Applying a treatment of electronic correlation effects beyond the DFT single-particle picture, namely in the form of local dynamic correlations within DFT+DMFT calculations, 
 we found a stable A-AFM ordered phase in the unstrained bulk LMO crystal structure.} Application of uniaxial strain results in a rather rare ferromagnetic insulating state, which is elusive to the single particle DFT formalism and can only be captured through many-body treatment of correlations within DMFT. In stark contrast, in case of tensile uniaxial strain generated by matching the shorter in plane lattice parameter to the longer, an antiferromagnetic insulating state is observed, similar to the case of unstrained LMO, albeit with a slightly larger band gap. Importantly, we also clarify that the reduction in JT distortion due to uniaxial compressive strain is a key ingredient for the emergence of the exotic ferromagnetic insulating state, along with the proper treatment of dynamic correlations. \textcolor{black}{Interestingly, our study shows that it is the flavour of
the strain (compressive/tensile) which has a major effect on the JT distortions and hence magnetism and not the effect of geometry/symmetry of matching to a square lattice.}
Our theoretical study can give rise to further experimental studies on the application of uniaxial strain, particularly in LMO, to validate the FM insulating state obtained for uniaxial compression. 

\section*{Acknowledgements}
This work has been funded by the Austrian Science Fund (FWF): Y746. Calculations have partly been performed on the cluster of TU Graz, and Vienna Scientific Cluster (VSC). 
\textcolor{black}{HB acknowledges generous support from Yusuf Hamied Department of Chemistry, University of Cambridge, and School of Science and Engineering, University of Dundee.}

 \bibliographystyle{elsarticle-num} 
 \bibliography{name.bib}

\providecommand{\noopsort}[1]{}\providecommand{\singleletter}[1]{#1}%
\begin{thebibliography}{10}
\expandafter\ifx\csname url\endcsname\relax
  \def\url#1{\texttt{#1}}\fi
\expandafter\ifx\csname urlprefix\endcsname\relax\def\urlprefix{URL }\fi
\expandafter\ifx\csname href\endcsname\relax
  \def\href#1#2{#2} \def\path#1{#1}\fi

\bibitem{HWANG2019100}
J.~Hwang, Z.~Feng, N.~Charles, X.~R. Wang, D.~Lee, K.~A. Stoerzinger, S.~Muy,
  R.~R. Rao, D.~Lee, R.~Jacobs, D.~Morgan, Y.~Shao-Horn,
  \href{https://www.sciencedirect.com/science/article/pii/S1369702119300185}{Tuning
  perovskite oxides by strain: Electronic structure, properties, and functions
  in (electro)catalysis and ferroelectricity}, Materials Today 31 (2019)
  100--118.
\newblock \href {https://doi.org/https://doi.org/10.1016/j.mattod.2019.03.014}
  {\path{doi:https://doi.org/10.1016/j.mattod.2019.03.014}}.
\newline\urlprefix\url{https://www.sciencedirect.com/science/article/pii/S1369702119300185}

\bibitem{franchini}
J.~He, C.~Franchini,
  \href{https://link.aps.org/doi/10.1103/PhysRevB.86.235117}{Screened hybrid
  functional applied to $3{d}^{0}\ensuremath{\rightarrow}3{d}^{8}$
  transition-metal perovskites {La$M$O$_{3}$} ({$M$} = {Sc}--{Cu}): Influence
  of the exchange mixing parameter on the structural, electronic, and magnetic
  properties}, Phys. Rev. B 86 (2012) 235117.
\newblock \href {https://doi.org/10.1103/PhysRevB.86.235117}
  {\path{doi:10.1103/PhysRevB.86.235117}}.
\newline\urlprefix\url{https://link.aps.org/doi/10.1103/PhysRevB.86.235117}

\bibitem{Ohtomo2004}
A.~Ohtomo, H.~Y. Hwang, \href{https://doi.org/10.1038/nature02308}{A
  high-mobility electron gas at the {LaAlO$_3$/SrTiO$_3$} heterointerface},
  Nature 427~(6973) (2004) 423--426.
\newblock \href {https://doi.org/10.1038/nature02308}
  {\path{doi:10.1038/nature02308}}.
\newline\urlprefix\url{https://doi.org/10.1038/nature02308}

\bibitem{hicks}
C.~W. Hicks, D.~O. Brodsky, E.~A. Yelland, A.~S. Gibbs, J.~A.~N. Bruin, M.~E.
  Barber, S.~D. Edkins, K.~Nishimura, S.~Yonezawa, Y.~Maeno, A.~P. Mackenzie,
  Strong increase of {$T_c$} of {Sr$_2$RuO$_4$} under both tensile and
  compressive strain, Science 344~(6181) (2014) 283--285.
\newblock \href {https://doi.org/10.1126/science.1248292}
  {\path{doi:10.1126/science.1248292}}.

\bibitem{satpathy}
S.~Satpathy, Z.~S. Popovi\ifmmode~\acute{c}\else \'{c}\fi{}, F.~R.
  Vukajlovi\ifmmode~\acute{c}\else \'{c}\fi{},
  \href{https://link.aps.org/doi/10.1103/PhysRevLett.76.960}{Electronic
  structure of the perovskite oxides:
  {${\mathrm{La}}_{1\ensuremath{-}\mathit{x}}{\mathrm{Ca}}_{\mathit{x}}{\mathrm{MnO}}_{3}$}},
  Phys. Rev. Lett. 76 (1996) 960--963.
\newblock \href {https://doi.org/10.1103/PhysRevLett.76.960}
  {\path{doi:10.1103/PhysRevLett.76.960}}.
\newline\urlprefix\url{https://link.aps.org/doi/10.1103/PhysRevLett.76.960}

\bibitem{zhang}
Z.~Liao, J.~Zhang,
  \href{https://www.mdpi.com/2076-3417/9/1/144}{Metal-to-insulator transition
  in ultrathin manganite heterostructures}, Applied Sciences 9~(1) (2019).
\newblock \href {https://doi.org/10.3390/app9010144}
  {\path{doi:10.3390/app9010144}}.
\newline\urlprefix\url{https://www.mdpi.com/2076-3417/9/1/144}

\bibitem{pavarini2010}
E.~Pavarini, E.~Koch,
  \href{https://link.aps.org/doi/10.1103/PhysRevLett.104.086402}{Origin of
  jahn-teller distortion and orbital order in ${\mathrm{lamno}}_{3}$}, Phys.
  Rev. Lett. 104 (2010) 086402.
\newblock \href {https://doi.org/10.1103/PhysRevLett.104.086402}
  {\path{doi:10.1103/PhysRevLett.104.086402}}.
\newline\urlprefix\url{https://link.aps.org/doi/10.1103/PhysRevLett.104.086402}

\bibitem{Lee_2013_PAPER}
J.~H. Lee, K.~T. Delaney, E.~Bousquet, N.~A. Spaldin, K.~M. Rabe,
  \href{https://link.aps.org/doi/10.1103/PhysRevB.88.174426}{Strong coupling of
  {Jahn-Teller} distortion to oxygen-octahedron rotation and functional
  properties in epitaxially strained orthorhombic {LaMnO3}}, Physical Review B
  88~(17) (2013) 174426.
\newblock \href {https://doi.org/10.1103/physrevb.88.174426}
  {\path{doi:10.1103/physrevb.88.174426}}.
\newline\urlprefix\url{https://link.aps.org/doi/10.1103/PhysRevB.88.174426}

\bibitem{gong}
Y.~S. Hou, H.~J. Xiang, X.~G. Gong,
  \href{https://link.aps.org/doi/10.1103/PhysRevB.89.064415}{Intrinsic
  insulating ferromagnetism in manganese oxide thin films}, Phys. Rev. B 89
  (2014) 064415.
\newblock \href {https://doi.org/10.1103/PhysRevB.89.064415}
  {\path{doi:10.1103/PhysRevB.89.064415}}.
\newline\urlprefix\url{https://link.aps.org/doi/10.1103/PhysRevB.89.064415}

\bibitem{lmo-sto-expt}
W.~S. Choi, D.~W. Jeong, S.~S.~A. Seo, Y.~S. Lee, T.~H. Kim, S.~Y. Jang, H.~N.
  Lee, K.~Myung-Whun,
  \href{https://link.aps.org/doi/10.1103/PhysRevB.83.195113}{Charge states and
  magnetic ordering in lamno${}_{3}$/srtio${}_{3}$ superlattices}, Phys. Rev. B
  83 (2011) 195113.
\newblock \href {https://doi.org/10.1103/PhysRevB.83.195113}
  {\path{doi:10.1103/PhysRevB.83.195113}}.
\newline\urlprefix\url{https://link.aps.org/doi/10.1103/PhysRevB.83.195113}

\bibitem{Rivero_2016_PAPERa}
P.~Rivero, V.~Meunier, W.~Shelton, Uniaxial pressure-induced half-metallic
  ferromagnetic phase transition in {LaMnO3}, Physical Review B 93~(9) (2016)
  094409.
\newblock \href {https://doi.org/10.1103/physrevb.93.094409}
  {\path{doi:10.1103/physrevb.93.094409}}.

\bibitem{zunger1}
J.~Varignon, M.~Bibes, A.~Zunger,
  \href{https://link.aps.org/doi/10.1103/PhysRevB.100.035119}{Mott gapping in
  $3d\phantom{\rule{0.16em}{0ex}}ab{\mathrm{o}}_{3}$ perovskites without
  mott-hubbard interelectronic repulsion energy u}, Phys. Rev. B 100 (2019)
  035119.
\newblock \href {https://doi.org/10.1103/PhysRevB.100.035119}
  {\path{doi:10.1103/PhysRevB.100.035119}}.
\newline\urlprefix\url{https://link.aps.org/doi/10.1103/PhysRevB.100.035119}

\bibitem{zunger2}
J.~Varignon, M.~Bibes, A.~Zunger,
  \href{https://link.aps.org/doi/10.1103/PhysRevResearch.1.033131}{Origins
  versus fingerprints of the jahn-teller effect in $d$-electron $ab{X}_{3}$
  perovskites}, Phys. Rev. Res. 1 (2019) 033131.
\newblock \href {https://doi.org/10.1103/PhysRevResearch.1.033131}
  {\path{doi:10.1103/PhysRevResearch.1.033131}}.
\newline\urlprefix\url{https://link.aps.org/doi/10.1103/PhysRevResearch.1.033131}

\bibitem{Varignon2019}
J.~Varignon, M.~Bibes, A.~Zunger,
  \href{https://doi.org/10.1038/s41467-019-09698-6}{Origin of band gaps in 3d
  perovskite oxides}, Nature Communications 10~(1) (2019) 1658.
\newblock \href {https://doi.org/10.1038/s41467-019-09698-6}
  {\path{doi:10.1038/s41467-019-09698-6}}.
\newline\urlprefix\url{https://doi.org/10.1038/s41467-019-09698-6}

\bibitem{banerjee2019}
H.~Banerjee, O.~Janson, K.~Held, T.~Saha-Dasgupta,
  \href{https://link.aps.org/doi/10.1103/PhysRevB.100.115143}{Electronic and
  magnetic state of {${\mathrm{LaMnO}}_{3}$} epitaxially strained on
  {${\mathrm{SrTiO}}_{3}$}: Effect of local correlation and nonlocal exchange},
  Phys. Rev. B 100 (2019) 115143.
\newblock \href {https://doi.org/10.1103/PhysRevB.100.115143}
  {\path{doi:10.1103/PhysRevB.100.115143}}.
\newline\urlprefix\url{https://link.aps.org/doi/10.1103/PhysRevB.100.115143}

\bibitem{banerjee-mplb}
H.~Banerjee, \href{https://doi.org/10.1142/S0217984920300069}{Understanding the
  role of exchange and correlations in complex oxides under strain and oxide
  heterostructures}, Modern Physics Letters B 34~(23) (2020) 2030006.
\newblock \href
  {http://arxiv.org/abs/https://doi.org/10.1142/S0217984920300069}
  {\path{arXiv:https://doi.org/10.1142/S0217984920300069}}, \href
  {https://doi.org/10.1142/S0217984920300069}
  {\path{doi:10.1142/S0217984920300069}}.
\newline\urlprefix\url{https://doi.org/10.1142/S0217984920300069}

\bibitem{banerjee2020}
H.~Banerjee, M.~Aichhorn,
  \href{https://link.aps.org/doi/10.1103/PhysRevB.101.241112}{Emergence of a
  ferromagnetic insulating state in
  {${\mathrm{LaMnO}}_{3}/{\mathrm{SrTiO}}_{3}$} heterostructures: Role of
  strong electronic correlations and strain}, Phys. Rev. B 101 (2020) 241112.
\newblock \href {https://doi.org/10.1103/PhysRevB.101.241112}
  {\path{doi:10.1103/PhysRevB.101.241112}}.
\newline\urlprefix\url{https://link.aps.org/doi/10.1103/PhysRevB.101.241112}

\bibitem{Kresse_1993_PAPER}
G.~Kresse, J.~Hafner, Ab initio molecular dynamics for liquid metals, Physical
  Review B 47~(1) (1993) 558--561.
\newblock \href {https://doi.org/10.1103/physrevb.47.558}
  {\path{doi:10.1103/physrevb.47.558}}.

\bibitem{Kresse_1996_PAPER}
G.~Kresse, J.~Furthmüller, Efficient iterative schemes for ab initio
  total-energy calculations using a plane-wave basis set, Physical Review B
  54~(16) (1996) 11169--11186.
\newblock \href {https://doi.org/10.1103/physrevb.54.11169}
  {\path{doi:10.1103/physrevb.54.11169}}.

\bibitem{Kresse_1996_PAPERa}
G.~Kresse, J.~Furthmüller, Efficiency of ab-initio total energy calculations
  for metals and semiconductors using a plane-wave basis set, Computational
  Materials Science 6~(1) (1996) 15--50.
\newblock \href {https://doi.org/10.1016/0927-0256(96)00008-0}
  {\path{doi:10.1016/0927-0256(96)00008-0}}.

\bibitem{Kresse_1999_PAPER}
G.~Kresse, D.~Joubert, From ultrasoft pseudopotentials to the projector
  augmented-wave method, Physical Review B 59~(3) (1999) 1758--1775.
\newblock \href {https://doi.org/10.1103/physrevb.59.1758}
  {\path{doi:10.1103/physrevb.59.1758}}.

\bibitem{Monkhorst_1976_PAPER}
H.~J. Monkhorst, J.~D. Pack, Special points for {Brillouin}-zone integrations,
  Physical Review B 13~(12) (1976) 5188--5192.
\newblock \href {https://doi.org/10.1103/physrevb.13.5188}
  {\path{doi:10.1103/physrevb.13.5188}}.

\bibitem{PBE_1996_PAPER}
J.~P. Perdew, K.~Burke, Y.~Wang, Generalized gradient approximation for the
  exchange-correlation hole of a many-electron system, Physical Review B
  54~(23) (1996) 16533--16539.
\newblock \href {https://doi.org/10.1103/physrevb.54.16533}
  {\path{doi:10.1103/physrevb.54.16533}}.

\bibitem{Blaha_2020_PAPER}
P.~Blaha, K.~Schwarz, F.~Tran, R.~Laskowski, G.~K.~H. Madsen, L.~D. Marks,
  {WIEN2k}: {An APW}+lo program for calculating the properties of solids, The
  Journal of Chemical Physics 152~(7) (2020) 074101.
\newblock \href {https://doi.org/10.1063/1.5143061}
  {\path{doi:10.1063/1.5143061}}.

\bibitem{Aichhorn_2009_PAPER}
M.~Aichhorn, L.~Pourovskii, V.~Vildosola, M.~Ferrero, O.~Parcollet, T.~Miyake,
  A.~Georges, S.~Biermann, Dynamical mean-field theory within an augmented
  plane-wave framework: Assessing electronic correlations in the iron pnictide
  {LaFeAsO}, Physical Review B 80~(8) (2009) 085101.
\newblock \href {https://doi.org/10.1103/physrevb.80.085101}
  {\path{doi:10.1103/physrevb.80.085101}}.

\bibitem{Aichhorn_2016_PAPER}
M.~Aichhorn, L.~Pourovskii, P.~Seth, V.~Vildosola, M.~Zingl, O.~E. Peil,
  X.~Deng, J.~Mravlje, G.~J. Kraberger, C.~Martins, M.~Ferrero, O.~Parcollet,
  {TRIQS}/{DFTTools}: A {TRIQS} application for ab initio calculations of
  correlated materials, Computer Physics Communications 204 (2016) 200--208.
\newblock \href {https://doi.org/10.1016/j.cpc.2016.03.014}
  {\path{doi:10.1016/j.cpc.2016.03.014}}.

\bibitem{Parcollet_2015_PAPER}
O.~Parcollet, M.~Ferrero, T.~Ayral, H.~Hafermann, I.~Krivenko, L.~Messio,
  P.~Seth,
  \href{http://www.sciencedirect.com/science/article/pii/S0010465515001666}{{TRIQS}:
  {A} toolbox for research on interacting quantum systems}, Computer Physics
  Communications 196 (2015) 398 -- 415.
\newblock \href {https://doi.org/https://doi.org/10.1016/j.cpc.2015.04.023}
  {\path{doi:https://doi.org/10.1016/j.cpc.2015.04.023}}.
\newline\urlprefix\url{http://www.sciencedirect.com/science/article/pii/S0010465515001666}

\bibitem{werner1}
P.~Werner, A.~Comanac, L.~de' Medici, M.~Troyer, A.~J. Millis,
  \href{https://link.aps.org/doi/10.1103/PhysRevLett.97.076405}{Continuous-time
  solver for quantum impurity models}, Phys. Rev. Lett. 97 (2006) 076405.
\newblock \href {https://doi.org/10.1103/PhysRevLett.97.076405}
  {\path{doi:10.1103/PhysRevLett.97.076405}}.
\newline\urlprefix\url{https://link.aps.org/doi/10.1103/PhysRevLett.97.076405}

\bibitem{Gull_2011_PAPER}
E.~Gull, A.~J. Millis, A.~I. Lichtenstein, A.~N. Rubtsov, M.~Troyer, P.~Werner,
  Continuous-time {Monte}~{Carlo} methods for quantum impurity models, Reviews
  of Modern Physics 83~(2) (2011) 349--404.
\newblock \href {https://doi.org/10.1103/revmodphys.83.349}
  {\path{doi:10.1103/revmodphys.83.349}}.

\bibitem{Seth_2016_PAPER}
P.~Seth, I.~Krivenko, M.~Ferrero, O.~Parcollet, {TRIQS}/{CTHYB}: A
  continuous-time quantum {Monte} {Carlo} hybridisation expansion solver for
  quantum impurity problems, Computer Physics Communications 200 (2016)
  274--284.
\newblock \href {https://doi.org/10.1016/j.cpc.2015.10.023}
  {\path{doi:10.1016/j.cpc.2015.10.023}}.

\bibitem{ct-seg}
We use the CTSEG implementation by T. Ayral, H. Hafermann, P. Delange, M.
  Ferrero and O. Parcollet, which is based on the TRIQS package.

\bibitem{Anisimov_1993_PAPER}
V.~I. Anisimov, I.~V. Solovyev, M.~A. Korotin, M.~T. Czy{\.{z}}yk, G.~A.
  Sawatzky, Density-functional theory and {NiO} photoemission spectra, Physical
  Review B 48~(23) (1993) 16929--16934.
\newblock \href {https://doi.org/10.1103/physrevb.48.16929}
  {\path{doi:10.1103/physrevb.48.16929}}.

\bibitem{maxent}
G.~J. Kraberger, R.~Triebl, M.~Zingl, M.~Aichhorn,
  \href{https://link.aps.org/doi/10.1103/PhysRevB.96.155128}{Maximum entropy
  formalism for the analytic continuation of matrix-valued {Green's}
  functions}, Phys. Rev. B 96 (2017) 155128.
\newblock \href {https://doi.org/10.1103/PhysRevB.96.155128}
  {\path{doi:10.1103/PhysRevB.96.155128}}.
\newline\urlprefix\url{https://link.aps.org/doi/10.1103/PhysRevB.96.155128}

\bibitem{Jain_2013_PAPER}
A.~Jain, S.~P. Ong, G.~Hautier, W.~Chen, W.~D. Richards, S.~Dacek, S.~Cholia,
  D.~Gunter, D.~Skinner, G.~Ceder, K.~A. Persson, {Commentary: The Materials
  Project: A} materials genome approach to accelerating materials innovation,
  {APL} Materials 1~(1) (2013) 011002.
\newblock \href {https://doi.org/10.1063/1.4812323}
  {\path{doi:10.1063/1.4812323}}.

\bibitem{LMO_MP_CELL}
mp~17554,
  \href{https://legacy.materialsproject.org/materials/mp-17554/#}{{LaMnO3}
  super cell from materials project database. mp-id 17554}, Lagacy Materials
  Project Database (Oct. 2020).
\newline\urlprefix\url{https://legacy.materialsproject.org/materials/mp-17554/#}

\bibitem{Dudarev_1998_PAPER}
S.~L. Dudarev, G.~A. Botton, S.~Y. Savrasov, C.~J. Humphreys, A.~P. Sutton,
  Electron-energy-loss spectra and the structural stability of nickel oxide:
  {An} {LSDA+U} study, Physical Review B 57~(3) (1998) 1505--1509.
\newblock \href {https://doi.org/10.1103/physrevb.57.1505}
  {\path{doi:10.1103/physrevb.57.1505}}.

\bibitem{Momma_2011_PAPER}
K.~Momma, F.~Izumi, {VESTA} for three-dimensional visualization of crystal,
  volumetric and morphology data, Journal of Applied Crystallography 44~(6)
  (2011) 1272--1276.
\newblock \href {https://doi.org/10.1107/s0021889811038970}
  {\path{doi:10.1107/s0021889811038970}}.

\bibitem{Rivero_2016_PAPER}
P.~Rivero, V.~Meunier, W.~Shelton, Electronic, structural, and magnetic
  properties of lamno3 phase transition at high temperature, Physical Review B
  93~(2) (2016) 024111.
\newblock \href {https://doi.org/10.1103/physrevb.93.024111}
  {\path{doi:10.1103/physrevb.93.024111}}.

\bibitem{ELEMANS1971238}
J.~B. Elemans, B.~V. Laar, K.~V.~D. Veen, B.~Loopstra,
  \href{https://www.sciencedirect.com/science/article/pii/002245967190034X}{The
  crystallographic and magnetic structures of la1-xbaxmn1-xmexo3 (me = mn or
  ti)}, Journal of Solid State Chemistry 3 (1971) 238--242.
\newblock \href {https://doi.org/https://doi.org/10.1016/0022-4596(71)90034-X}
  {\path{doi:https://doi.org/10.1016/0022-4596(71)90034-X}}.
\newline\urlprefix\url{https://www.sciencedirect.com/science/article/pii/002245967190034X}

\bibitem{Elemans_1971_PAPER}
E.~J., L.~B., The crystallographic and magnetic structures of
  {La$_{1-x}$Ba$_{x}$Mn$_{1-x}$Me$_{x}$O$_{3}$} ({Me} = {Mn} or {Ti}), Journal
  of Solid State Chemistry 3~(2) (1971) 238--242.
\newblock \href {https://doi.org/10.1016/0022-4596(71)90034-x}
  {\path{doi:10.1016/0022-4596(71)90034-x}}.

\bibitem{georges_rmp}
A.~Georges, G.~Kotliar, W.~Krauth, M.~J. Rozenberg,
  \href{https://link.aps.org/doi/10.1103/RevModPhys.68.13}{Dynamical mean-field
  theory of strongly correlated fermion systems and the limit of infinite
  dimensions}, Rev. Mod. Phys. 68 (1996) 13--125.
\newblock \href {https://doi.org/10.1103/RevModPhys.68.13}
  {\path{doi:10.1103/RevModPhys.68.13}}.
\newline\urlprefix\url{https://link.aps.org/doi/10.1103/RevModPhys.68.13}

\bibitem{banerjee2022}
H.~Banerjee, H.~Schnait, M.~Aichhorn, T.~Saha-Dasgupta,
  \href{https://link.aps.org/doi/10.1103/PhysRevB.105.235106}{Effect of
  geometry on magnetism of {Hund's} metals: Case study of
  {${\mathrm{BaRuO}}_{3}$}}, Phys. Rev. B 105 (2022) 235106.
\newblock \href {https://doi.org/10.1103/PhysRevB.105.235106}
  {\path{doi:10.1103/PhysRevB.105.235106}}.
\newline\urlprefix\url{https://link.aps.org/doi/10.1103/PhysRevB.105.235106}

\bibitem{banerjee2023}
H.~Banerjee, C.~P. Grey, A.~J. Morris,
  \href{https://link.aps.org/doi/10.1103/PhysRevB.108.165124}{Importance of
  electronic correlations in exploring the exotic phase diagram of layered
  ${\mathrm{li}}_{x}{\mathrm{mno}}_{2}$}, Phys. Rev. B 108 (2023) 165124.
\newblock \href {https://doi.org/10.1103/PhysRevB.108.165124}
  {\path{doi:10.1103/PhysRevB.108.165124}}.
\newline\urlprefix\url{https://link.aps.org/doi/10.1103/PhysRevB.108.165124}

\end{thebibliography}





\end{document}